\documentclass[twocolumn,tightenlines,nofootinbib,prd,amsmath,amssymb]{revtex4-1}

\usepackage{amsmath, amsfonts, amsthm, amssymb, graphicx}
\usepackage{epstopdf}
\usepackage{slashed}
 \usepackage{graphicx,epsfig}
\usepackage{amsmath}
\usepackage {amssymb}
\usepackage{hyperref}
\usepackage[T1]{fontenc}
\usepackage{fancyhdr}
\usepackage{anysize}

\usepackage{graphicx}
\usepackage{amsmath}
\usepackage{amsfonts}
\usepackage{mathrsfs}
\usepackage{epsfig}
\usepackage{lineno}
\usepackage{color}
\usepackage{hyperref}

\begin{document}

\title{The Virasoro Gibbs state and BTZ black holes}

\author{Alan Garbarz}
\email{alan@df.uba.ar}
\affiliation{Departamento de F\'isica Juan Jos\'e Giambiagi, FCEyN UBA and IFIBA CONICET-UBA,
Facultad de Ciencias Exactas y Naturales, Ciudad Universitaria, Pabell\'on I, 1428 Buenos Aires, Argentina
 }


\date{\today}

\begin{abstract}
We show that the  Virasoro Gibbs state accurately describes the thermodynamics of BTZ black holes at large temperatures and in the strong-coupling regime $c<1$. We first give a simple heuristic argument by showing that at high temperatures and arbitrary $c$, the energies are much larger than Planck mass. Then we give a detailed analysis of the quantum fluctuations of the Gibbs state on unitary irreducible representations of Virasoro group and explicitly show that they go to zero in the high-temperature limit by means of representation theory results. This implies the state has a sensible thermodynamic limit which actually holds for any $c$. Finally, the matching with BTZ thermodynamics for $c<1$ is obtained by using the known asymptotic behaviour of the characters of the Virasoro discrete series. This result supports the idea that minimal models could describe quantum gravity at strong coupling. We make no use of the Euclidean path integral nor assume modular invariance of the partition functions, although we comment on them along the way. 
\end{abstract}

\maketitle

\section{Introduction}
Three-dimensional Einstein gravity with a negative cosmological constant and its quantization has been the subject of great study in the last years (see \cite{KP} and references therein). On one hand, the classical theory is intimately related to Chern-Simons theory \cite{Witten88,AT86} and the phase space of solutions contains the well-known BTZ black holes \cite{BTZ}. Even more, giving the fact that all solutions are locally diffeomorphic to AdS$_3$, the covariant phase space of solutions is tractable, depending heavily on the chosen  boundary conditions. The usual ones are called Brown-Henneaux boundary conditions \cite{BrownHenneaux}, which give rise to a set of asymptotic symmetries Vect$(S^1)\times$Vect$(S^1)$ and imply that the asymptotic solutions of \cite{Banadossolutions} transform between each other through (the coadjoint action of) two copies of the Virasoro group with central charge $c=3\ell/2G$. 

The complete classification of the asymptotic solutions in terms of coadjoint orbits of Virasoro group was performed in \cite{GL,BO}. Among many observations, it was noted that each BTZ belongs to a different Virasoro orbit, all being geometrically Diff$(S^1)/S^1$. AdS$_3$ belongs to the only orbit of the form Diff$(S^1)/$PSL$(2,\mathbb{R})$. Every Virasoro coadjoint orbit is an infinite-dimensional symplectic manifold and for those just mentioned the energy associated to $L_0$  is bounded from below.  In general, classical observables are labelled by elements of Vect$(S^1)\times$Vect$(S^1)$, and this is why we use, in what follows, results coming from the study of the so-called conformal nets on $S^1$ (chiral local CFTs, see for example \cite{Carpi}).

On the other hand, quantization can be approached at least from two angles, by means of the Euclidean path integral or by finding unitary representations of the classical symmetries on some Hilbert space. We adopt the latter. We do not however take advantage of the classical Chern-Simons formulation as a starting point. We instead assume that all the symmetries we need are those that come from the asymptotic transformations given by the Virasoro group. This assumption should be dropped if one attempts to perform the quantization of the full classical theory taking into account what the geometries could look like far enough from the boundary.

The Virasoro group and its representations have been long studied in the literature (see for example \cite{Carpi} and \cite{S} for complementary discussions on this). The unitary irreducible positive-energy representations were first obtained in \cite{GW} and recently classified in \cite{NS} as vector bundles over Diff$(S^1)/S^1$ (without requiring irreducibility). 

Since we consider only the phase space of asymptotic solutions, and we intend to be as minimalistic as possible, in order to describe the quantum version of a BTZ black hole we consider as Hilbert space (the completion of) a single irreducible unitary highest-weight representation of Virasoro group. This can be thought as the quantization of a single Virasoro orbit Diff$(S^1)/S^1$ (which is not the one corresponding to that black hole). It is expected that a quantization of the complete space of solutions such as that of \cite{KS} contains in some way the Virasoro group representations here considered, which is actually the case in the quantization of \cite{KP} (see  Section V).

It is worth mentioning that, on an irreducible representation, it is already known that there is not enough degeneracy of descendants to account for the entropy for large energies and large $c$: the $L_0$ eigenvalue $h+n$ has degeneracy $p(n)$, the partition of $n$, which grows with $n$ independently of $c$, so it cannot reproduce the BTZ entropy which does depend on $c$. So usually one drops the irreducibility requirement and considers having a distribution of primary states together with its descendants. In addition, the constraint of modular invariance of the partition function is imposed (an Euclidean path-integral approach in this direction is \cite{MaloneyWitten}). This permits the use of the Cardy formula to count degeneracies and thus the BTZ entropy \cite{Strominger}.  

We, on the contrary, explore a different setting: we keep the irreducibility of the Hilbert space and then we can only expect to reproduce the BTZ thermodynamics for $c<1$, which is in principle a regime where one would not expect to match the semiclassical behaviour of black holes. Despite of this, it has been assumed in \cite{Gaberdiel} that for $c<1$ the path-integral decomposes in exactly the same way as in the weak-coupling $c>>1$ case. We shall not discuss this assumption here, nevertheless the present work can be considered as complementary to \cite{Gaberdiel}, in the sense that, from a representation theory angle, we explain how the Gibbs state reproduces the BTZ thermodynamics when $c<1$. Note that, as opposed to \cite{Gaberdiel}, working on irreducible representations discards modular invariance of the partition functions. However, our results further justify why each of the characters appearing in the modular partition functions of \cite{Gaberdiel} can be understood as coming from BTZ contributions: those characters are the (grand) canonical partition functions considered here \footnote{The cases where $h>c/24$ in \cite{Gaberdiel} are related to BTZ contributions. This should be revisited in the light of the present work, since $h$ is not the energy of a BTZ. We discuss this further in Section IV.}.

Let us now state the results: first we show in Section III that, for any $c$, quantum fluctuations of hermitian operators in the Gibbs state vanish \footnote{Actually for hermitian combinations of $L_n$ with $n\neq 0$ one can see the fluctuations are non-zero but bounded, which is the same as with the microcanonical density operator.} at high temperatures (`thermodynamic limit' from now on), proving that the canonical ensemble matches the microcanonical one and that there is a sensible thermodynamic limit. Then, in Section IV, we see that in this limit the BTZ thermodynamics are reproduced when $c<1$. This condition on the central charge is at the opposite end of the traditional  semi-classical approximation $c>>1$, so  let us discuss it.

For the moment, until we discuss in detail quantum fluctuations, we can give an heuristic argument of why large temperatures are an \textit{alternative} thermodynamic limit for arbitrary $c$: the Virasoro algebra implies that the characteristic quantum energy is $\hbar/\ell$. Also, the dimensionless quantum central charge becomes $c=3\ell/2G\hbar$. A sensible way of interpreting the semi-classical limit  $c>>1$ is that the classical energies of order $G^{-1}$ are much larger than  the typical quantum energy $\hbar/\ell$. Analogously, but now taking the large-temperature limit, classical energies are again much larger than $\hbar/\ell$, even in the strong-coupling regime $c\sim 1$. This is because once a quantum state is considered it introduces a new energy scale, $\langle L_0\rangle$, and if this state describes a BTZ black hole, we can compare its energy $M$ to $\hbar/\ell$ which gives
\begin{equation}
\frac{M}{\hbar/\ell} = \frac{c}{12} \frac{r_+^2}{\ell^2}. 
\end{equation} 
Thus large classical energies (large mean values of $L_0$) can take place for any fixed $c$ as long as we consider big (hot) enough BTZ black holes. In short, the thermodynamic limit $r_+>>\ell$ where quantum fluctuations disappear is consistent with the physically sensible classical region of large energies, despite the value of $c$.

\section{Thermodynamics of BTZ black holes}

Here we remind the reader of the thermodynamic relations of the BTZ black hole \cite{BTZ}. The spacetime can be parameterized by two positive quantities $r_+$ and $r_-$ with length units given by $\ell$:
\begin{eqnarray}
ds^ 2&=&-\frac{(r^2-r_+^2)(r^2-r_-^2)}{\ell^2 r^2}dt^2+\\
&&\frac{\ell^2 r^2dr^2}{(r^2-r_+^2)(r^2-r_-^2)}+r^2\left( d\phi-\frac{r_+ r_-}{\ell r^2}dt\right)^2.\nonumber
\end{eqnarray}
The mass $M$ and angular momentum $J$ are given by 
\begin{equation}\label{massandangularmomentum}
M=\frac{r_+^2+r_-^2}{8G\ell^2}, \quad J=\frac{2r_+ r_-}{8G\ell},
\end{equation}
where an ad hoc energy scale $G^{-1}$ is introduced (there is no Newtonian limit). In addition, it has been pointed  out,  first in \cite{BTZ} and then in many places in the literature, that this black hole satisfies the first law  of thermodynamics, $dM=T dS + \Omega dJ $, with temperature and entropy given by 
\begin{equation}\label{tempandentropy}
\beta_{\text{BTZ}}^{-1}=T=\frac{\hbar}{2\pi\ell^2}\frac{r_+^2-r_-^2}{r_+},\quad S=\frac{2\pi r_+}{4G\hbar}. 
\end{equation}
Of course, this is to be taken as a semi-classical estimation, in the light of the (classical) black hole r\^ole  in presence of a quantum field. In the first law, $\Omega=r_-/\ell r_+$ is the angular potential. The whole thermodynamic behaviour can be summarized in a \textit{semi-classical} grand canonical partition function \cite{Banados},
\begin{equation}
Z_{\text{BTZ}}(\beta_+,\beta_-)=e^{-\beta_{\text{BTZ}} F}=Z_{1/2}(\beta_+)\times Z_{1/2}(\beta_-),
\end{equation} 
where, 
\begin{equation}\label{BTZpartitionfunction}
Z_{1/2}(\beta)=\exp\left(c\frac{\pi^2}{6}\frac{\ell}{\hbar\beta}\right)
\end{equation}
with $\beta_\pm=\beta_{\text{BTZ}}(1\pm \ell\Omega)$ and  $c=3\ell/2G\hbar$ (the division by $\hbar$ is convenient for later purposes). It is important to emphasize once again that $Z_{\text{BTZ}}$ condensates the entire thermodynamics of the black hole. In fact, it is evident that only one copy (\ref{BTZpartitionfunction}) is enough and the complete picture is just the product of two copies (with different temperatures). The Bekenstein-Hawking entropy, for example, is obtained by taking the sum of entropies of each copy.  We show in the rest of the letter that the partition function of a discrete series representation of Virasoro group has precisely the form (\ref{BTZpartitionfunction}) in the thermodynamic limit.

\section{The Virasoro Gibbs state}\label{Gibbsstate}

The main point of this section is to show that a Gibbs state on certain Virasoro module has a well defined thermodynamic limit at high temperatures. As far as we know, this has not been shown before in the literature, and here we make use of results on conformal nets on $S^1$.

We consider a unitary irreducible representation of the Virasoro group on a Hilbert space $\mathcal{H}_{h,c}$, which is the completion of an irreducible unitary Verma module $V_{h,c}$ \cite{GW}. This implies that either $c\geq 1$ and $h\geq 0$, or we have the discrete series representations which are labelled by certain discrete values of $1/2\leq c<1$ and $0\le h$ \cite{S}. In this section everything is dimensionless.  

The canonical partition function with Hamiltonian given by $L_0$ is,
\begin{equation}\label{canonicalpartitionfunction}
Z(\beta):=\text{Tr} \exp(-\beta L_0)=\frac{e^{-\beta h} }{\prod\limits_{n\geq 1} (1-e^{-\beta n} )}, 
\end{equation}
where the second equality only holds for $c\geq 1$. For the case $c<1$ more involved expressions appear (see \cite{KR}), and we only need a few properties of them. Equation (\ref{canonicalpartitionfunction}) is the character of the one parameter group generated by $L_0$. Usually $\beta$ is replaced by a complex number $2\pi i\tau$ and Im $2\pi\tau=\beta>0$. 

The state we want to study is the Gibbs state at inverse temperature $\beta$. It is the only KMS-state with respect to rotations on the circle, so there is no alternative \cite{LongoTanimoto}: if one assumes that the representation is irreducible with fixed $h$ and that the correct quantum state $\rho$ describing a black hole must be thermal, then the only possibility is,
\begin{equation}\label{rho}
\rho=\frac{e^{-\beta L_0}}{Z}.
\end{equation} 
Actually, this is the case for all $c$ in the discrete series and also for $c\geq 1$. 

First of all, let us see that this density operator entails a thermodynamic behaviour at large temperatures by proving that the relative fluctuation of the energy tends to zero in this limit. A short typical computation in statistical mechanics gives
\begin{equation}
\frac{\langle L_0^2 \rangle - \langle L_0 \rangle^2}{\langle L_0 \rangle^2}=\frac{\partial_\beta^2 \log Z }{(\partial_\beta \log Z)^2}
\end{equation}
If $\partial_\beta\log Z$ goes as $\beta^{-2}$ for $\beta\rightarrow 0$, then the relative dispersion in the energy goes to zero at high temperatures and the statistical description coincides with thermodynamics \footnote{We do not consider the $\beta\rightarrow\infty$ limit since it makes the Gibbs state tend exponentially fast to the primary vector which has zero entropy.}. 

It has been proven in \cite{KL} that the logarithm of the  character $Z$ has the following asymptotic expansion for any modular conformal net on $S^1$,
\begin{equation}\label{Longoasymptotic}
\log Z \rightarrow \frac{ c\pi^2}{6} \frac{1}{\beta} + \text{corrections},\quad \beta\rightarrow 0^+.  
\end{equation}
From now on we will always mean asymptotic expansion when writing that certain function approaches another one. The condition that the CFT should be modular in order for (\ref{Longoasymptotic}) to hold restricts the values of $c$ to be $c<1$ if the quantum theory is to have only Virasoro operators \footnote{A Virasoro net on $S^1$ is not modular when $c \geq 1$. We thank R. Longo and Y. Tanimoto for clarifications on this issue.}. So we conclude that,  since $\partial_\beta \log Z$ goes as $\beta^{-2}$ at least when $c<1$, the density state (\ref{rho}) accurately describes  a thermodynamic system in the high-temperature limit \footnote{We are using Theorem 4.1 in \cite{Olver} to justify differentiation of an asymptotic expansion, since it can be shown $\log Z$ meets the requirements.}.

Now we turn to the case with $c \geq 1$, where we do not have modularity of the CFT and so there is no result like the one used for $c<1$. The approach is more or less known and consists of exploiting the quasi-modularity of the divergent piece of $\partial_\beta\log Z$ to compute its asymptotic form. From (\ref{canonicalpartitionfunction}) we have
\begin{equation}\label{partiallogZ}
-\partial_\beta \log Z =h+\sum_{n\geq 1} \frac{n q^n}{1-q^n}=h+\sum_{n\geq 1} \sigma_1(n) q^n,
\end{equation}
where $\sigma_1(n)$ is the sum of the divisors of $n$, $q=e^{2\pi i \tau}$, and the last equality comes from performing the Fourier expansion. The number $h$ is independent of $\beta$ so the divergent part comes from  $f(q):=\sum \sigma_1(n) q^n$, which is related to the normalized Eisenstein series  $G_2(\tau)$  \cite{Serre},
$$G_2(\tau)=\frac{\pi^2}{3}\left(1-24 f(q)\right)$$
This quasi-modular form satisfies
\begin{equation}
\tau^{-2}G_2(-1/\tau)=G_2(\tau)-\frac{2\pi i}{\tau}
\end{equation} 
from where one can obtain, with $\tau=i\beta/2\pi$ and $G_2(i\infty)=\pi^2/3$,
\begin{equation}\label{fasymptotics}
f(q)\rightarrow\frac{\pi^2}{6\beta^2} + \text{corrections},\quad \beta\rightarrow 0^+
\end{equation}
Then, again, relative fluctuations of the energy go to zero for high temperatures. So far we have shown the Virasoro Gibbs state entails a thermodynamic limit for large temperatures and any $c$. Actually, to complete the picture, it can be shown that all other hermitian combinations of Virasoro operators have zero-mean value, and their fluctuations tend exponentially to the fluctuations of descendants with the same $\langle L_0\rangle$. In this sense, the Gibbs state tends to the microcanonical denisty state (thermodynamic limit) and the energy $\langle L_0 \rangle$ becomes large (compared to the typical gap of one unit).

We are interested in a conformal theory with two independent copies of Virasoro operators on the circle. So everything is easily repeated taking into account that the (grand canonical) partition function is $Z_{2d}=Z\times \bar Z$, and the Gibbs state of inverse temperature $\beta_{2d}$ is the product of two copies: $\rho_{2d}=\rho \otimes \bar\rho$ at inverse temperatures $\beta$ and $\bar\beta$ ($\beta_{2d}$ is half $\beta+\bar \beta$). For example, the relative energy fluctuation is,
\begin{equation}
\frac{\Delta( L_0 + \bar L_0)^2 }{\langle L_0 + 
\bar L_0\rangle^2} = \frac{\partial_\beta^2 \log Z + \partial_{\bar{\beta}}^2 \log \bar Z}{(\partial_\beta \log Z + \partial_{\bar{\beta}} \log \bar Z)^2}.
\end{equation}
Thus, if for both copies $\partial_{\beta}\log Z \sim \beta^{-2}$ for small $\beta$, then again fluctuations tend to zero. However, we continue working with only one copy to keep the notation simpler, and where necessary we will make explicit the changes in order to account for the two copies.

From the Gibbs state and the partition function one can for example compute the equation of state and the von Neumann entropy:
\begin{equation}
\langle L_0 \rangle =\frac{1}{Z}\text{Tr} \left(L_0 e^{-\beta L_0}\right)=-\partial_\beta \log Z,
\end{equation}
\begin{equation}\label{entropy}
S=-\text{Tr}\,\rho\log\rho=\log Z+\beta\langle L_0 \rangle 
\end{equation}
These statistical quantities sharply describe their classical analogues when the relative fluctuations are small. We have shown that this is the case for the Virasoro Gibbs state at high temperatures.

\section{Recovering the thermodynamics of BTZ black holes}

Now we turn our attention to explain how the Gibbs state accurately reproduces the thermodynamics of large-temperature BTZ geometries when $c<1$. The point we want to make is that in this large-temperature limit the partition function  of a discrete series Virasoro representation is exactly as in (\ref{BTZpartitionfunction}). In such limit, the Gibbs state (\ref{rho}) should describe the same thermodynamic behaviour as a black hole of large temperature $\beta_{\text{BTZ}}^{-1}$ with mass $M$ and angular momentum $J$. We have already seen that for any $c$, $\partial_\beta \log Z\propto \beta^{-2}$ at high temperatures, so  the Gibbs state entails a thermodynamic limit and moreover it seems to have the appropriate dependence with $\beta$. It is left to see that the proportionality constant is the one in (\ref{BTZpartitionfunction}). This is precisely the case for the discrete series, but before showing this, a few preliminary words about units are in order. 

In order to use the results of the previous section, we should make the thermodynamic quantities dimensionless. The canonical Virasoro observables have units of $\hbar$, so one should consider $L_0/\ell$ as the quantum Hamiltonian with energy units. This can also be explained by the fact that $L_0/\ell$ is (half of) the time-translation generator in the gravity theory. The Virasoro algebra then implies that the fundamental gaps between eigenvalues of energy are $\hbar/\ell$. In addition, the dimensionless version of the central element in the quantum theory is $c=3\ell/2G\hbar$. Actually, as usual, dividing the quantum Lie bracket relations by $\hbar$ makes every Virasoro operator dimensionless and the results of the previous section hold without change. As for the temperature of our Gibbs state, it should be measured against the typical quantum energy, so we multiply  the dimensionful $\beta$ times $\hbar/\ell$. The limit of high temperatures is $\beta\rightarrow\beta_{BTZ}$ with $\beta_{BTZ}<<\ell/\hbar$. As explained in the introduction, this translates into large black holes: $r_+\pm r_->>\ell$. From now on we work with dimensionless quantities.

Let us start with $c<1$. From the discussion above we consider $\beta$ in the high temperature  limit: $\beta\rightarrow\beta_-$ and $\bar\beta\rightarrow\beta_+$ (recall that $\beta_\pm$ are defined after (\ref{BTZpartitionfunction}) and are now dimensionless). Thus, each Virasoro partition function (\ref{Longoasymptotic})  matches the (half) BTZ partition function (\ref{BTZpartitionfunction}). This means all BTZ thermodynamic properties are recovered precisely in the thermodynamic limit of the Virasoro  Gibbs state. This is our main result. For example the entropy (\ref{entropy}) is obtained as,
\begin{equation}
S= c\frac{\pi^2}{3\beta_+}+c\frac{\pi^2}{3\beta_-}=\frac{2\pi r_+}{4G\hbar}.
\end{equation}  
As a cautionary note, it is worth mentioning that the few allowed values of $h$ and $\bar h$ of the discrete series do not influence the large-temperature limit (it only depends on $c$). For any of them, the Gibbs state describes a BTZ specified by the value that $\beta$ approaches to, namely the inverse temperature $\beta_{\text{BTZ}}$. In particular, for a BTZ with mass $M$ and angular momentum $J$,  
\begin{equation}\label{MJrelations}
\hbar\langle L_0+\bar L_0\rangle\rightarrow\ell M  ,\quad  \hbar\langle L_0 - \bar L_0\rangle\rightarrow J ,
\end{equation}
at large temperatures. This means that the primary sate of weight $h$ does not describe the BTZ thermodynamics. Instead, the Gibbs state mimics the behaviour of a mixture of fixed-energy pure states with a large scaling dimension given by $c\pi^2/6\beta^{2}_{\text{BTZ}}$. In particular this implies that all the (discrete series) characters in the partition functions of \cite{Gaberdiel} can be thought as coming from contributions of BTZs at large temperatures (not only those characters where $h>c/24$).

Now let us move to the case $c\geq 1$, where $h$ is no longer constrained (should be non-negative though) and in addition since there is no modularity of the partition functions, (\ref{Longoasymptotic}) does not hold. Moreover, we know from (\ref{partiallogZ}) and (\ref{fasymptotics}) that at large temperatures $-\partial_\beta \log Z \rightarrow h + \pi^2/6\beta^2$. This implies that 
\begin{equation}\label{logZ}
\log Z \rightarrow -h\beta +  \pi^2/6\beta	,\quad \beta\rightarrow 0^+,
\end{equation}
with the possible addition of a constant which ultimately would shift the entropy, and we set it to zero. We see right away that if $h$ is big enough so that $h\beta$ is of the same order as $1/\beta$, the $\beta$ dependence of $\log Z$ in \ref{logZ} is not $\beta^{-1}$ and so we do not recover the semi-classical partition function of a BTZ black hole (\ref{BTZpartitionfunction}).  If $h$ is not big enough then $\log Z \rightarrow \pi^2/6\beta$, but againg it does not give the right thermodynamics for $c>1$, since the $c$ factor is missing (compare to \ref{Longoasymptotic}).

\section{Final remarks}

The unique thermal state on a unitary positive-energy irreducible representation of the Virasoro group has a well-defined thermodynamic limit at large temperatures (namely, relative fluctuations of $L_0$ go to zero or are bounded for the other hermitian operators). Surprisingly, in the strong-coupling regime $c<1$, this state accurately reproduces the thermodynamics of a BTZ black hole of large temperature,  with mass and angular momentum given by (\ref{MJrelations}). In particular, these results show that the characters appearing in the physical partition functions of \cite{Gaberdiel} are always related to some BTZ, even when $h<c/24$. This is so because the black hole is not described by the primary state of weight $h$ (this one has zero entropy). It is actually described by the Gibbs state at large temperature, with an  energy mean value matching the mass of the BTZ, which is always larger than $c/24$ independently of $h$.

On the other hand, for $c>1$, the BTZ thermodynamics is not reproduced. Simply put, $\log Z$ should go as in (\ref{Longoasymptotic}), namely proportional to the central charge. The reason this does not happen could be that there are not enough quantum observables in order to further constrain the representations, and thus their characters, so that modular transformations similar to those of the characters when $c<1$ occur and the central charge appears in the expression of $\log Z$. However, it would be nice that this enlargement of the algebra of operators was not just an educated guess, but that it came from the quantization of a larger set of classical observables. If the new larger algebra is modular with same Brown-Henneaux central charge, then by (\ref{Longoasymptotic}) the Gibbs state (if it is well-defined) reproduces the correct thermodynamics at large temperatures.  

A possible larger set of classical observables appears in the very nice work of Krasnov and Scarinci \cite{KS}, where the solution space of AdS$_3$ gravity is studied with great generality and detail. Notably, they make a precise parametrization of solutions which are defined not only asymptotically, but in the whole manifold. The group of symplectomorphisms of their ``universal phase space'' could be the one to be quantized and hopefully would enlarge the algebra of operators in such a way that accounts for the thermodynamics of black hole solutions. These classical symmetries have not been studied in full yet, as far as we know. An interesting work in this direction is \cite{KP}, where a K\"ahler quantization of the phase space of \cite{KS} was performed for $c>>1$, resulting in a Hilbert space in some sense too large. This fact led the authors of \cite{KP} to postulate a projection, motivated by path-integral arguments, onto a smaller Hilbert space. Further study of this proposal and of the classical symmetries of \cite{KS} will most likely give a better understanding of quantum AdS$_3$ and its thermal states.  

\section*{Acknowledgments}

This work was supported by UBA and CONICET. We thank G. Giribet, M. Leston, G. P\'erez-Nadal and J. Zanelli for many fruitful conversations and for suggesting valuable changes to the preliminary versions of the present letter. We also thank R. Longo and Y. Tanimoto for correspondence regarding their works, and in particular we acknowledge many discussions with the latter on local conformal nets and their relation to black hole physics.

\end{document}